\documentclass[pra,showpacs,superscriptaddress,twocolumn,10pt]{revtex4-1}
\usepackage[latin1]{inputenc}
\usepackage[colorlinks=true,linkcolor=blue,urlcolor=blue,citecolor=blue,pdfusetitle]{hyperref}
\usepackage[cm]{fullpage}
\usepackage{graphicx}	
\usepackage[english]{babel}
\usepackage{amsmath}
\usepackage{amssymb}
\usepackage{cancel}
\usepackage{natbib}
\usepackage{comment}
\usepackage{color}
\usepackage[normalem]{ulem}

\newcommand{\rrefs} {Refs.~}

\newcommand{\Tr}{\text{Tr}}

\makeatletter
	\renewcommand{\maketag@@@}[1]{\hbox{\m@th\normalsize\normalfont#1}}
\makeatother
\begin{document}
\author{Nadia Milazzo}
\affiliation{Institut f\"ur theoretische Physik, Universit\"at T\"ubingen, 72076 T\"ubingen, Germany and LPTMS, UMR 8626, CNRS, Univ. Paris-Sud, Universit\'e Paris-Saclay, 91405 Orsay, France   }
\author{Salvatore Lorenzo}
\affiliation{Dipartimento di Fisica e Chimica, Universit\`a  degli Studi di Palermo, via Archirafi 36, I-90123 Palermo, Italy}
\author{Mauro Paternostro}
\affiliation{Centre for Theoretical Atomic, Molecular, and Optical Physics, School of Mathematics and Physics, Queen's University, Belfast BT7 1NN, United Kingdom}
\author{G. Massimo Palma}
\affiliation{NEST, Istituto Nanoscienze-CNR and Dipartimento di Fisica e Chimica, Universit\`a  degli Studi di Palermo, via Archirafi 36, I-90123 Palermo, Italy}

\begin{abstract}
Quantum Darwinism attempts to explain the emergence of objective reality of the state of a quantum system in terms of redundant information about the system acquired by independent non interacting  fragments of the environment. The consideration of interacting environmental elements gives rise to a rich phenomenology, including the occurrence of non-Markovian features, whose effects on objectification {\it a' la} quantum Darwinism needs to be fully understood. We study a model of local interaction between a simple quantum system and a multi-mode environment that allows for a clear investigation of the interplay between information trapping and propagation in the environment and the emergence of quantum Darwinism. We provide strong evidence of the correlation between non-Markovianity and quantum Darwinism in such a model, thus providing strong evidence of a potential link between such fundamental phenomena. 

\end{abstract}


\title{The role of information back-flow in the emergence of quantum Darwinism}

\date{\today}

\maketitle


The quantum Darwinism paradigm is one of the most recent and convincing attempt to explain the emergence of objective reality out of superpositions of quantum states (for a review see~\cite{Zurek1}). In this framework, the first key mechanism responsible of the transition from quantum to classical is the coupling of the system with an environment which acquires information about the state of the system with respect the so called pointer states, namely  the eigenstates of the observable which is coupled with the environment \cite{Zurek2,Zurek2bis,Zurek2ter}.  
If the system is in a specific pointer state it is left undisturbed by such coupling, if however the system is in a coherent superposition of pointer states it gets entangled with the environment. 
An external observer who can access the environment can therefore acquire informations on the state of the system, leading to its objective existence. 

The second key ingredient at the basis of quantum Darwinism is the particular structure of the environmental states which get entangled with the pointer state. The basic idea is that in a real scenario the environment degrees of freedom are not traced out but rather accessed by different observers. The assumption is that information on the state of the system is redundantly encoded in multiple, independent fragments of the environment, which we assume to consist of a large set of non interacting units. External observers can read the information on the system contained in separate, locally accessible fragments of the environment, with each fragment containing the same information on the system, this leading to objective reality of the system state~\cite{Zurek3,Zurek4,Zurek5}. 
This is what naturally happens when an initial coherent superposition of system pointer states $|\Psi\rangle_S = \sum_{k=1}^n \psi_k |\pi_k\rangle_S$ evolves into a joint system-environment state with a branching structure  
\begin{equation}
|\Psi_{SE}\rangle = \sum_{k=1}^n \psi_k |\pi_k\rangle_S\bigotimes^M_{j=1}|\eta_k\rangle_j,
\end{equation}
where the information about the system state $|\pi_k\rangle_S$ is imprinted into multiple copies of environmental states $|\eta_k\rangle$, thus becoming accessible to individual, distinct observers, that access separate fragments of the environment. 

The phenomenology of quantum Darwinism has attracted a robust body of work, recently~\cite{ZwolakPRL2009, ZwolakPRA2010, ZurekNJP2012, ZurekSciRep2013, ZwolakPRL2014, ZurekSciRep2016, ZurekPRA2017, PianiNatComms2015, BalaneskovicaEPJD2015, MendlerEPJD2016, GarrawayPRA2017, Zambrini, Zambrini2,ZwolakArXiv}, while the first attempts at its experimental assessment have been reported~\cite{Brunner2008, Burke2010,Ciampini2018, Chen2018}. Yet, the fundamental mechanism for its emergence, and the features that characterize it are yet to be fully understood. On one hand, the relation with interesting alternative formulations for the emergence of objective reality through the formalism of quantum spectrum broadcasting structures needs clarifying~\cite{HorodeckiPRA2015, KorbiczPRL2014,LePRL2018}. On the other hand, it appears that quantum correlations have a significant influence on the qualification of quantum Darwinism~\cite{LePRA2018}. Such interplay deserves a complete understanding in light of the relevance that, say, quantum entanglement has for the characterization of the quantum-to-classical transition. Finally, and most relevantly for the work reported in this paper, the possible influence that memory effects in the open-system dynamics of a quantum information carrier have on the emergence of objective reality have been the focus of controversies. 
While recent studies have suggested the detrimental role of non-Markovianity for the manifestation of quantum Darwinism~\cite{Zambrini,Zambrini2}, the relation between objectification and spectrum broadcasting structures appears to be loose~\cite{Lampo}.

Quantum non-Markovianity is a core issue in a program for the grasping of the foundations of quantum Darwinism. In recent years quantum information theory has helped in understanding the physical meaning and in providing the mathematical tools to characterise quantum non-Markovian dynamics. In particular, a number of theoretical measures of the degree of quantum non-Markovianity of an open dynamics have been put forward~\cite{review1, review2, review3, review4, BreuerLP09, RivasHP10,LorenzoPP13}. Such measures have been used to identify the regions in the parameters space corresponding to Markovian and non-Markovian  dynamics for a variety of environmental models \cite{model1, model2, model3, model4, model5, model6, model7, model8, model9}. All such measures capture the idea that non-Markovianity is linked to a back-flow of quantum information from the environment to the system. 

Such a link could be key in understaanding the process of objectification that is at the core of Darwinism. In this paper we contribute to such an understanding by exploring the links between information back-flow and the emergence of quantum Darwinism in a physically relevant scenario that is rather different from the configurations addressed so far. We consider a two-level system locally coupled to a single harmonic oscillator that is part of a one-dimensional interacting harmonic lattice that embodies the environment. This is distinct from the typical assumption of a system being collectively coupled to the elements of the environment~\cite{Zambrini,Zambrini2}. Such a differencec is not to be underestimated: In this situation, in fact, we expect information on the state of the system not to be copied onto separate fragments of the environment. Rather, quantum information about the system would flow through the environmental fragments via their mutual interaction. We show that when the system shows a Markovian evolution it shows also a darwinistic behaviour while Darwinism disappears when the system dynamics is non-Markovian. We also explore how both effects are linked with the directionality of quantum information flow from the system to the environment. 

The remainder of the paper is structured as follows. In Sec.~\ref{model} we introduce our model, solve the reduced dynamics of the system of interest, which turns out to be a time-dependent dephasing one, and quantify the corresponding degree of non-Markovianity of such dynamics. Sec.~\ref{relation} is devoted to the phenomenology of quantum Darwinism as non-Markovian effects settle in into the dynamics of the system. In particular, in Sec.~\ref{propagation} we link such phenomenology to the features of information flow across the environmental lattice, thus providing a clear physical assessment for the onset of Darwinism and non-Markovianity. Sec.~\ref{conc} reports our conclusions.

\section{the model and its degree of non  Markovianity}
\label{model}

\begin{figure}[t]
\includegraphics[width=\columnwidth]{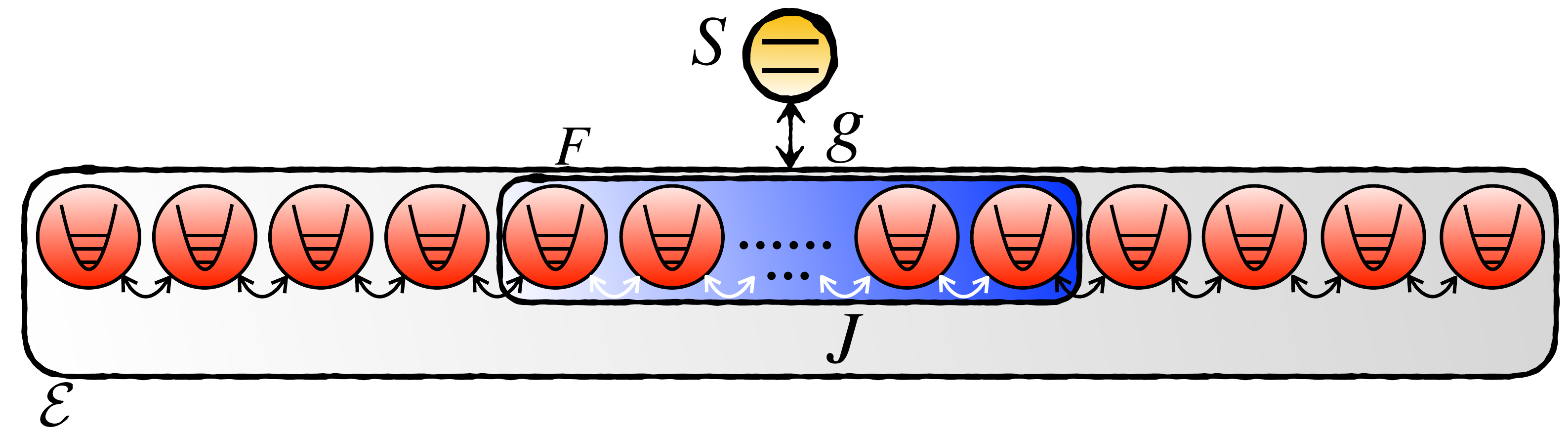}
\caption{A single two-level $S$ is locally coupled (at a strength $g$) to a harmonic oscillator that is part of a linear array of interacting oscillators ${\cal E}$. The inter-oscillator coupling rate is $J$. We study the mutual information that increasingly large fragments $F$ of the array share with system $S$. The fragments $F$ will all consist of sections centred at the qubit position.}
\label{array}
\end{figure}

We consider an exactly soluble model in which we drop the assumption of independent sub-environments. Specifically, we consider an environment ${\cal E}$ consisting of a one-dimensional array of $N$ linearly coupled harmonic oscillators. On the other hand, the system $S$ is embodied by a two-level system that is locally coupled to one of such oscillators [cf. Fig.~\ref{array}]. The Hamiltonian of the model (written in units such that $\hbar=1$) reads
\begin{equation}
\hat{H}=\frac{\omega_0}{2} \hat\sigma^S_z+\omega\sum_j\hat a^{\dagger}_j \hat a_j+J\sum_{j}(\hat a^{\dagger}_j \hat a_{j+1}+ h.c.)+g\sigma_z(\hat a^{\dagger}_0+\hat a_0),
\end{equation}
\label{h}

where $\omega_0 $ is the energy splitting between the states of $S$, $\hat \sigma^S_z$ is its $z$-Pauli matrix, $\omega $ is the frequency of the $j^\text{th}$ local oscillators with bosonic creation (annihilation) operator $ a^{\dagger}_j$ ($a_j $) and $ j\in\lbrace-{(N-1)}{/2},{(N-1)}/{2}\rbrace $. The bosonic intra-environment coupling rate is $J$. We assume that $S$ is locally coupled at a rate $g$ to the central oscillator, which has label $j=0$. 

Due to the coupling with the central harmonic oscillator, the information about the state of $S$ propagates along the environment ${\cal E}$, whose sections act as interacting environment fragments which can be individually accessed.
With such geometry, the way in which the environmental fragments acquire information about the state of the system is radically different from the non-interacting or the star-shaped scenario: quantum information must now propagate along the array and is not acquired simultaneously by the environmental fragments as in the case of a star geometry where the system is simultaneously coupled to $N$ (possibly interacting) sub-environments. We will show how the efficiency with which quantum information flows along the array affects the emergence of quantum Darwinism and determines non-Markovian effects in the reduced dynamics of $S$. In particular, we will show that the threshold in our parameter space at which we have an onset of non-Markovianity is exactly the same as the one at which quantum Darwinism breaks down. 

The key feature of our choice of environmental model is the possibility to switch from a local-oscillator picture to a one based on normal modes. The exactly solvable nature of the environmental Hamiltonian also simplify the establishment of a clear relationship between the information that is locally accessible and the information flux across ${\cal E}$. 

In terms of normal modes $\hat{b}_k$ of the environment, the Hamiltonian reads
\begin{equation}
\hat{H}=\frac{\omega_0}{2} \hat\sigma^S_z+\sum_k \Omega_k \hat{b}_k^{\dagger}\hat{b}_k+\frac{g}{\sqrt{N}} \hat\sigma_z\sum_k (\hat{b}_k^{\dagger}+\hat{b}_k),
\label{hd}
\end{equation}
where $\hat{b}_k^{\dagger}=\sum_j e^{ikj}\hat{a}^{\dagger}_j/\sqrt{N}$ is the creation operator of mode $k={2\pi n}/{N}$ ($ n=-\frac{N-1}{2},...,\frac{N-1}{2}$) and the frequency $\Omega_k$ of the $k^\text{th}$  normal mode is determined by the dispersion relation  
$\Omega_k=\omega+\omega_k=\omega+2 J \cos k$.

In the interaction picture with respect to the free-energy terms  $\frac{\omega_0}{2} \hat\sigma^S_z+\sum_k \Omega_k \hat{b}_k^{\dagger}\hat{b}_k$, the time evolution operator takes the form of a collection of conditional displacement operators, one per normal mode, dependent on the state of the two-level system~\cite{Palma}. Explicitly, we have
\begin{equation}
\hat{\cal{U}}_I(t) = e^{-i\phi}e^{-i\sigma_z\sum_k (\beta_k(t)\hat{b}^{\dagger}_k-\beta^*_k(t)\hat{b}_k)},
\label{U(t)}
\end{equation}
where $ \phi $ is an irrelevant global phase factor (arising from time ordering) and $\beta_k(t)=({g}/{\sqrt{N}})({1-e^{i\Omega_k t}})/{\Omega_k}$~\cite{Cirone}. Such propagator leads to decoherence, with the system states $ \lbrace \vert 0\rangle_S,\vert 1\rangle_S\rbrace $ as pointer states~\cite{Zurek2}. A linear superposition of the system's states such as $c_0\vert0\rangle_S+c_1\vert1\rangle_S$ -- with the environmental modes prepared in their vacuum state -- evolves into the entangled state
\begin{equation}
\vert\psi(t)\rangle_{S{\cal E}}=c_0\vert 0\rangle_S{\bigotimes}_k\vert -\beta_k(t)\rangle+c_1\vert 1\rangle_S{\bigotimes}_k\vert \beta_k(t)\rangle
\label{ps}
\end{equation}
where $\vert \beta_k(t)\rangle$ is a coherent state of mode $k$. Such dynamics leads to decoherence of the reduced density operator $\rho_S(t)$ of the system, with a decoherence function 
$e^{-\Gamma(t)} = \prod_k\langle\beta_k(t)\vert -\beta_k(t)\rangle$. An explicit calculation leads to~\cite{Palma}
\begin{equation}
\Gamma(t)=\frac{4g^2}{N}\sum_k  \frac{1-\cos\Omega_k t }{\Omega^2_k}.
\end{equation}
Such reduced dynamics can be ascribed to the time-local master equation~\cite{review3}
\begin{equation}
\dot\rho_S(t)=\gamma(t)[\sigma_z\rho_S(t)\sigma_z-\rho_S(t)],
\end{equation}
where $\gamma(t) $ is a {time-dependent decoherence rate} related to $\Gamma(t)$ as 
\begin{equation}
\gamma(t)=\frac{\dot\Gamma(t)}{2}=\frac{2g^2}{N}\sum_k \frac{\sin \Omega_k t}{\Omega_k}.
\end{equation} 
\begin{figure}[htbp]
\includegraphics[width=\columnwidth]{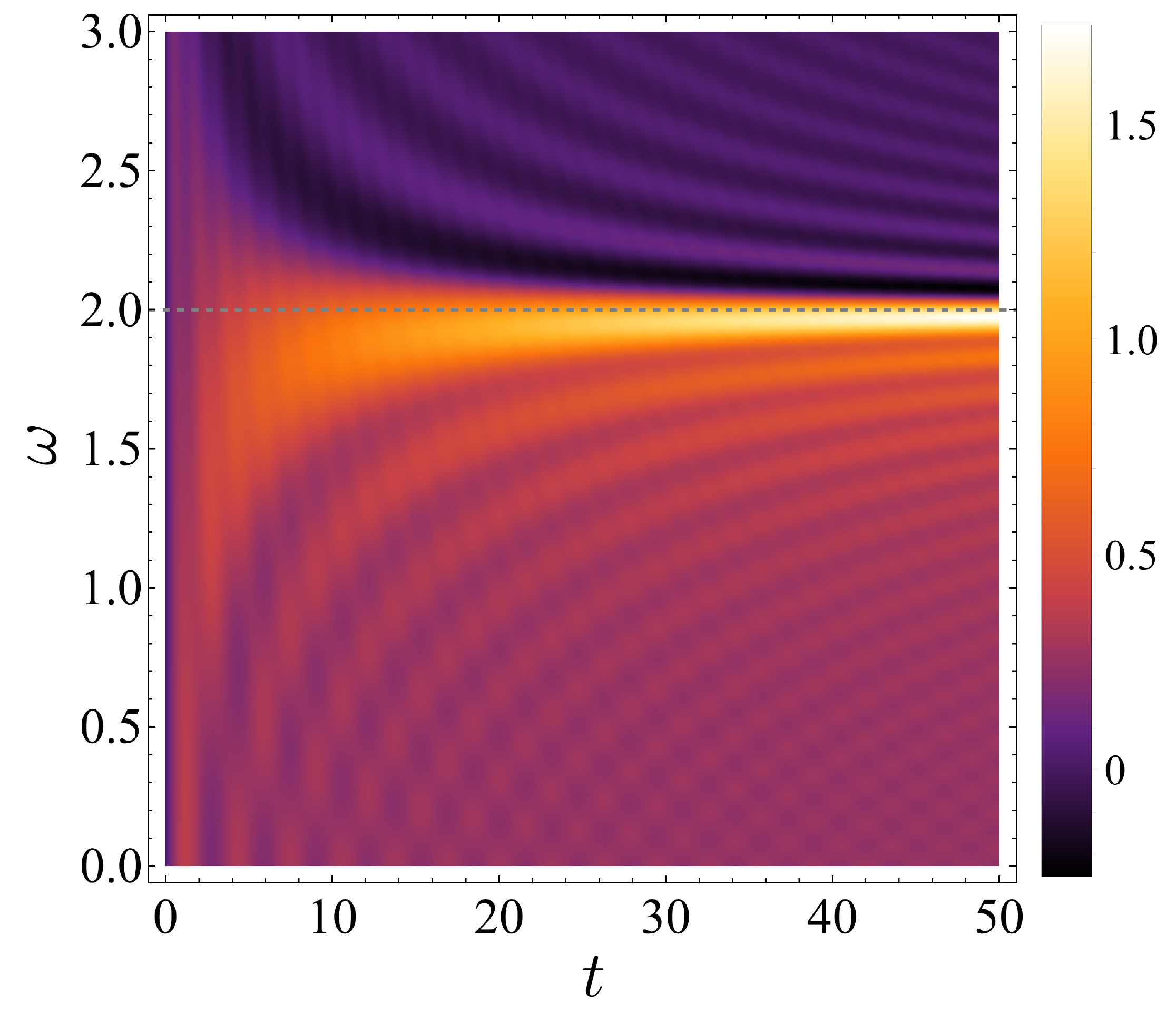}
     \caption{ The time dependent decay rate $\gamma(t)$ for $ \omega\in[0,3]$ (in units of $J$). A  transition between Markovian and non-Markovian dynamics emerges at $ \omega=2 $, 
     where $\gamma (t)$ reaches its maximum before assuming negative values  for $ \omega/J>2 $.}
     \label{densplot}
\end{figure}
The dependence of $\gamma(t)$ on $\omega$ has a strong influence on the degree of non-Markovianity of the system dynamics. Indeed while, in general, different non-Markovianity measures are associated with different partitions of the  parameter space characterising the open dynamics of the system, this is not the case for pure decoherence, where the measures introduced in \rrefs\cite{BreuerLP09, RivasHP10,LorenzoPP13}, which are relevant instances of informative tools for the characterization of non-Markovianity, lead to the same simple criterion for the occurrence of non-Markovian behaviour: pure dephasing of a qubit is Markovian (non-Markovian) iff $\gamma(t)\geq 0$ ($\gamma(t)< 0$). As shown in Fig.~\ref{densplot} a sharp transition in the sign of $\gamma(t)$ occurs for $\omega/J =2$. At such threshold value, $\gamma (t)$ reaches its maximum before turning negative for $\omega/J> 2$. 

\begin{figure}[htbp]
\includegraphics[width=\columnwidth]{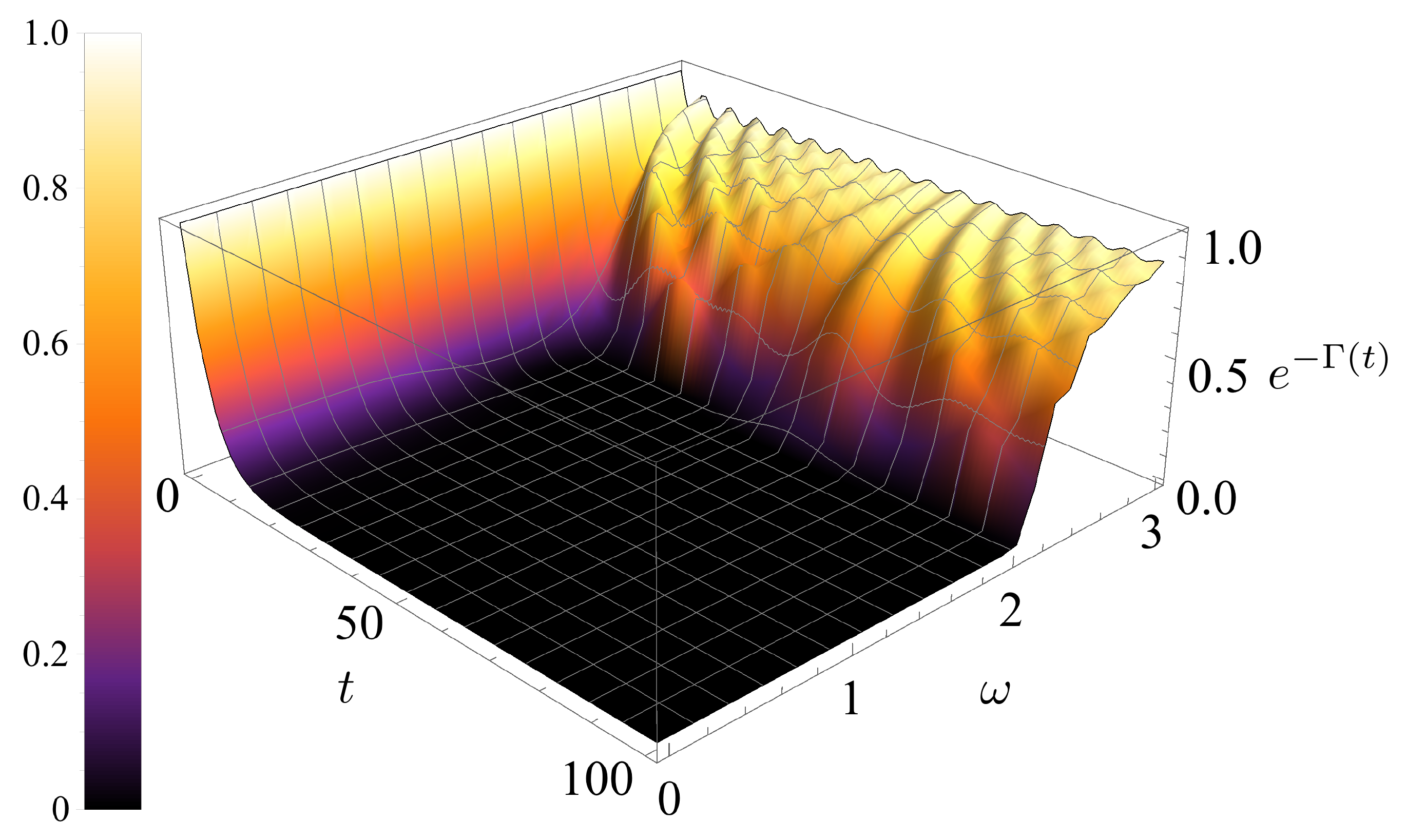}
\caption{Decoherence of a single qubit interacting with an environment of $ N=201 $ oscillators versus the evolution time \textit{t} and the local-oscillator frequency  $ \omega $ for $\textit{g} = 0.5 $ (all parameters in units of $J$). The change in the behavior of the decoherence function $e^{-\Gamma(t)}$ is clearly observed at $ \omega/J=2 $.}
\label{gamma}
\end{figure}

A strong deviation from pure monotonic dephasing in the region of non-Markovianity is evident in Fig.~\ref{gamma}, where $ e^{-\Gamma(t)} $ is plotted against the evolution time and the frequency $ \omega $ (in units of $J$). Again, a drastic change emerges for $ \omega/J>2 $ as a back-flow of information from the environment to the system occurs, rendering the time evolution of $S$ non-Markovian.

\section{Quantum Darwinism and non-Markovianity}
\label{relation}

The signature of Darwinism is the presence of a redundancy plateau in the so called \textit{Partial Information Plots} (PIP), i.e. the plot of the mutual information $I(S:F)$ shared by the system $S$ and the fragment $F$ of the environment ${\cal E}$ accessible by the observers, against the size of $F$ itself. In order to be quantitative, let us consider the reduced joint density operator of the system plus the fragment $\rho_{SF} = \Tr_{R}|\Psi_{S{\cal E}}\rangle \langle \Psi_{S{\cal E}}|$, where the trace is on all the elements of the environment except those belonging to the fragment \textit{F}. We thus call $R$ the part of the array that is traced out, so that ${\cal E}=F+R$. We have 

\begin{equation}
I(S:F) = S(\rho_S) + S(\rho_F) - S(\rho_{SF}),
\end{equation}

where $S(\rho ) = -\Tr(\rho \ln \rho)$ is the von Neumann entropy of state $\rho$. 

In the Darwinistic scenario, a PIP exhibits a typical redundant profile~\cite{ Zurek1}: $I(S:F)$ rapidly increases for small values of the dimension $f$ of the fraction being considered and then reaches a {plateau} at $ S(\rho_S)$. This entails the \textit{classical} plateau: when this amount of information on the state of $S$ is gained, further observations of other sub-environments (i.e. larger values of $f$) simply confirm what is already known about the system. The plateau is a characteristic ``footprint" of quantum Darwinism: all the fragments contain the same information about the system; the information obtained is objective, since many observers agree about the outcomes.

We now show how the environment of interacting oscillators gains and stores locally redundant information about the  system, leading to typical Darwinistic PIPs. To evaluate the quantum mutual information between the system $S$ and growing fractions of the environment it is necessary to go back from the the normal modes description to the local-oscillator one. This allows us to follow the dynamics and to evaluate the entropy of the reduced density operator of fragments consisting of finite sections of the oscillators array. 
The time evolution operator in Eq.~\eqref{U(t)} takes the form of a tensor product of conditional displacement operators each acting on a single bosonic normal mode $k$. Such structure is retained also when the time evolution operator is expressed in terms of local harmonic $j$ oscillators as
\begin{equation}
\hat{\cal{U}} (t)={\bigotimes}_k \hat{\cal{D}}_c(\beta_k)={\bigotimes}_j \hat{\cal{D}}_c(\alpha_j),
\end{equation}
where  
\begin{equation} 
\hat{\cal{D}}_c(\eta) = \exp\{\sigma_z(\eta\hat{\mu}^{\dagger}-\eta^*\hat{\mu})\}
\end{equation}
is the displacement operator with amplitude  $\eta=\alpha_j=\sum_k \beta_k e^{ikj}/{\sqrt{N}}$ ($\eta=\beta_k$) for $\hat\mu=\hat a_j$ ($\hat\mu=\hat b_k$).
As already mentioned, we study fragments $F$ of ${\cal E}$ consisting of sections centred at the oscillator labelled as $j=0$ and coupled to the system [cf. Fig.~\ref{array}]. However, it is worth mentioning that different choices of fragment arrangements leads to results consistent with what will be discussed in the following.  

Without  loss of generality, we assume as initial state of the system the balanced superposition $|\psi (0)\rangle_S = (|0\rangle_S + |1\rangle_S)/\sqrt{2}$ and the harmonic oscillator array in its ground state.
Needless to say, as long as the initial state of the harmonic array is pure, the state of the system-environment compound $ \rho_{S{\cal E}}(t) $ remains pure at all times and the von Neumann entropy of $\rho_\textit{SF}$ can be quantified through the one of its complementary part R. The corresponding mutual information is thus $I(S:F)=S(\rho_S)+S(\rho_F)-S(\rho_R)$. 

\begin{figure}
\includegraphics[width=\columnwidth]{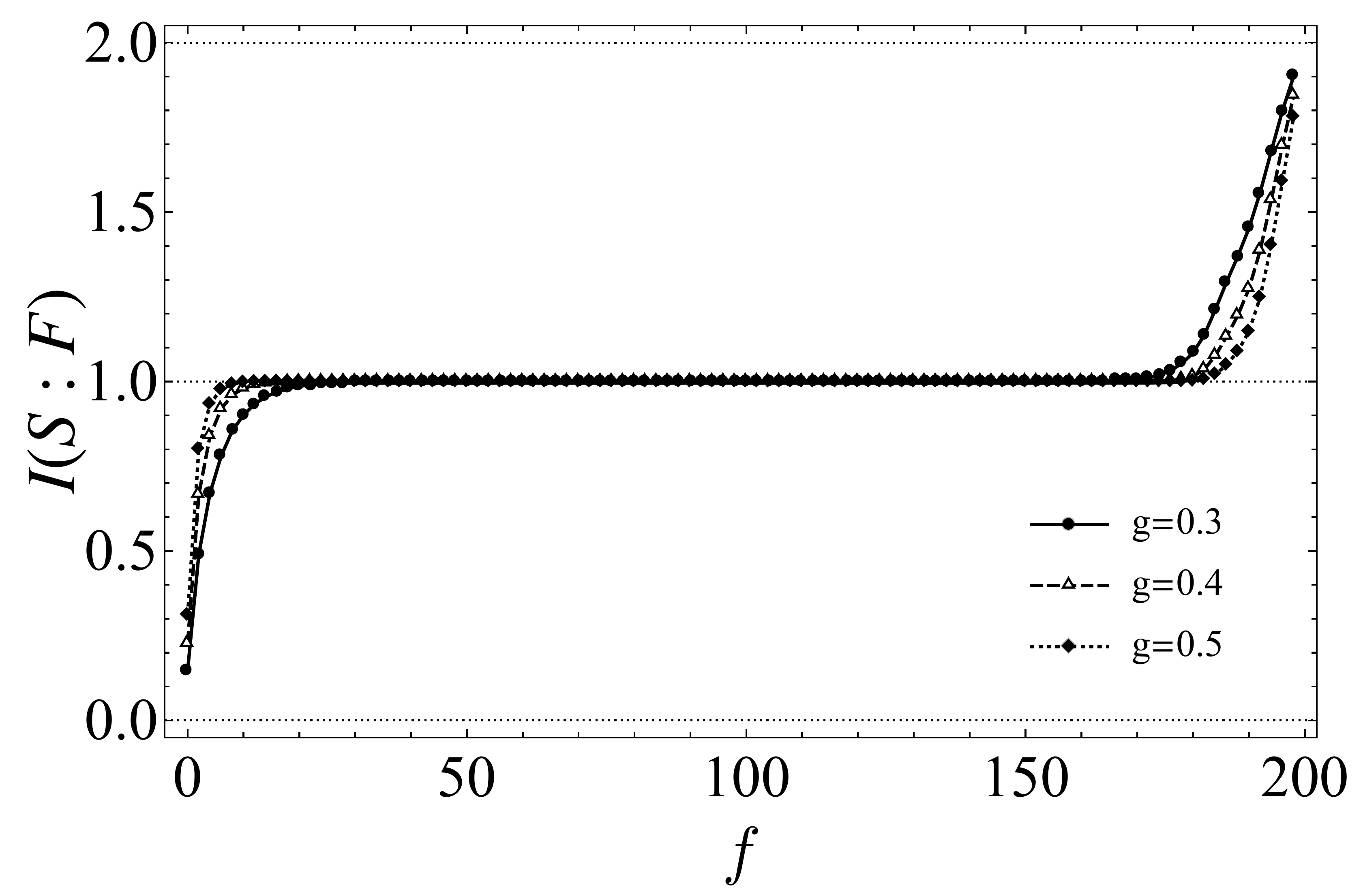}
\caption{PIPs showing $ I(S:F)$ against the number $f$ of elements in the considered fraction $F$ for an array of $N=201$ oscillators. The three different curves correspond to different values of \textit{g} (in units of $J$), for $ \omega/J=0.5 $. All the curves show a redundant behaviour, which is a characteristic feature of quantum Darwinism. By increasing the coupling constant \textit{g}, the $ I(S:F)$ curve show sharper onset of the plateau, which is a signature of higher degrees of redundancy. }
\label{darwin}
\end{figure}

We start by analyzing the emergence of Darwinism in the Markovian regime, i.e. when $ \omega<2 J $. In Fig.~\ref{darwin} we show PIPs for times long enough for the perturbation induced by the coupling with the system to reach the boundaries of the array, $ \omega/J=0.5 $ and  various choices of the coupling constant $g$, as a function of the number $f$ of elements in fragment $F$. The reported results should be taken as typical for the situation studied here. Remarkably, they all exhibit redundant behaviour: $I(S:F)$ rapidly increases at small fractions of the environment, then reaches a plateau at $ S(\rho_S) =1 $. When $ f\sim N $, i.e. the entire environment is accessed, $I(S:F)$ again  increases sharply and approaches $ 2S(\rho_S)$.  
Therefore, in the Markovian regime the presence of interactions between environmental fragments is not an obstacle to the emergence of quantum Darwinism and we still observe a redundant information encoding about the pointer observable $ \hat\sigma^S_z $ in the environment.

\begin{figure}
\includegraphics[width=\columnwidth]{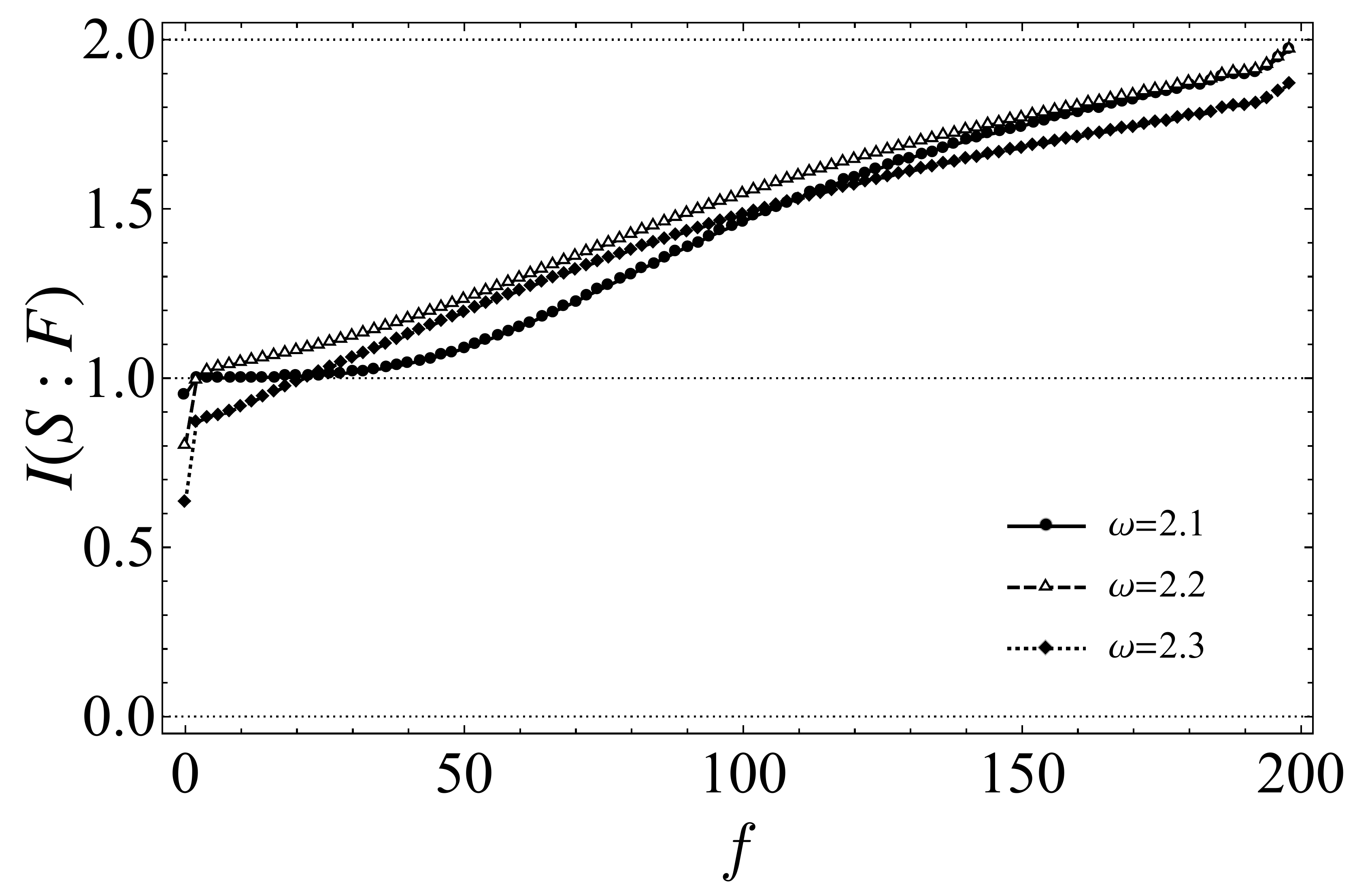}
\caption{PIPs similar to the plot reported in Fig.~\ref{darwin}. The different curves correspond to different values of $ \omega $, all greater than 2 (in units of $J$). The redundant encoding of information about \textit{S} is clearly lost: quantum Darwinism is not emergent, and the amount of the information grows with \textit{f}.}
\label{miomega}
\end{figure}

Above the threshold  $ \omega=2J$, we instead observe the loss of  Darwinistic behavior, as shown in Fig.~\ref{miomega}. There is no more a redundancy plateau, and adding fractions of the environment means increasing the amount of information about the system.
Darwinism and Markovianity indeed appear as parallel effects: when the time evolution is Markovian, that is $ \gamma(t)\geq 0$, the interacting environmental fragments still store redundant information about the pointer observable of the system. When the two-level system dynamics acquires non-Markovian features ($ \gamma(t)<0 $), quantum Darwinism features disappear, and the quantum mutual information between the system and different fractions of the environment grows approximately linearly.

 \section{information propagation along the array}
 \label{propagation}
 
\begin{figure*}[t]
{\bf (a)}\hskip7cm{\bf (b)}
\includegraphics[width=\columnwidth]{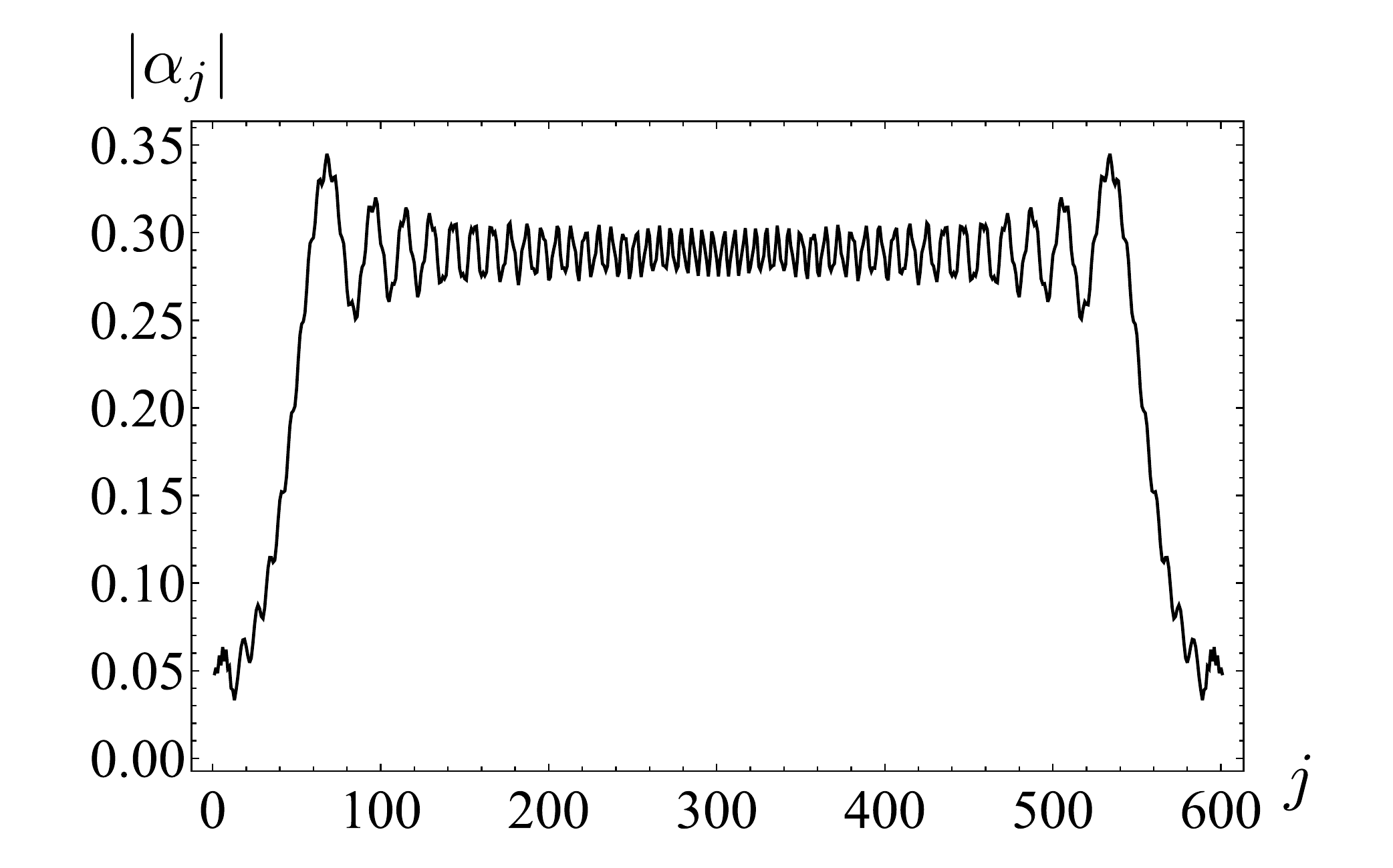}\includegraphics[width=\columnwidth]{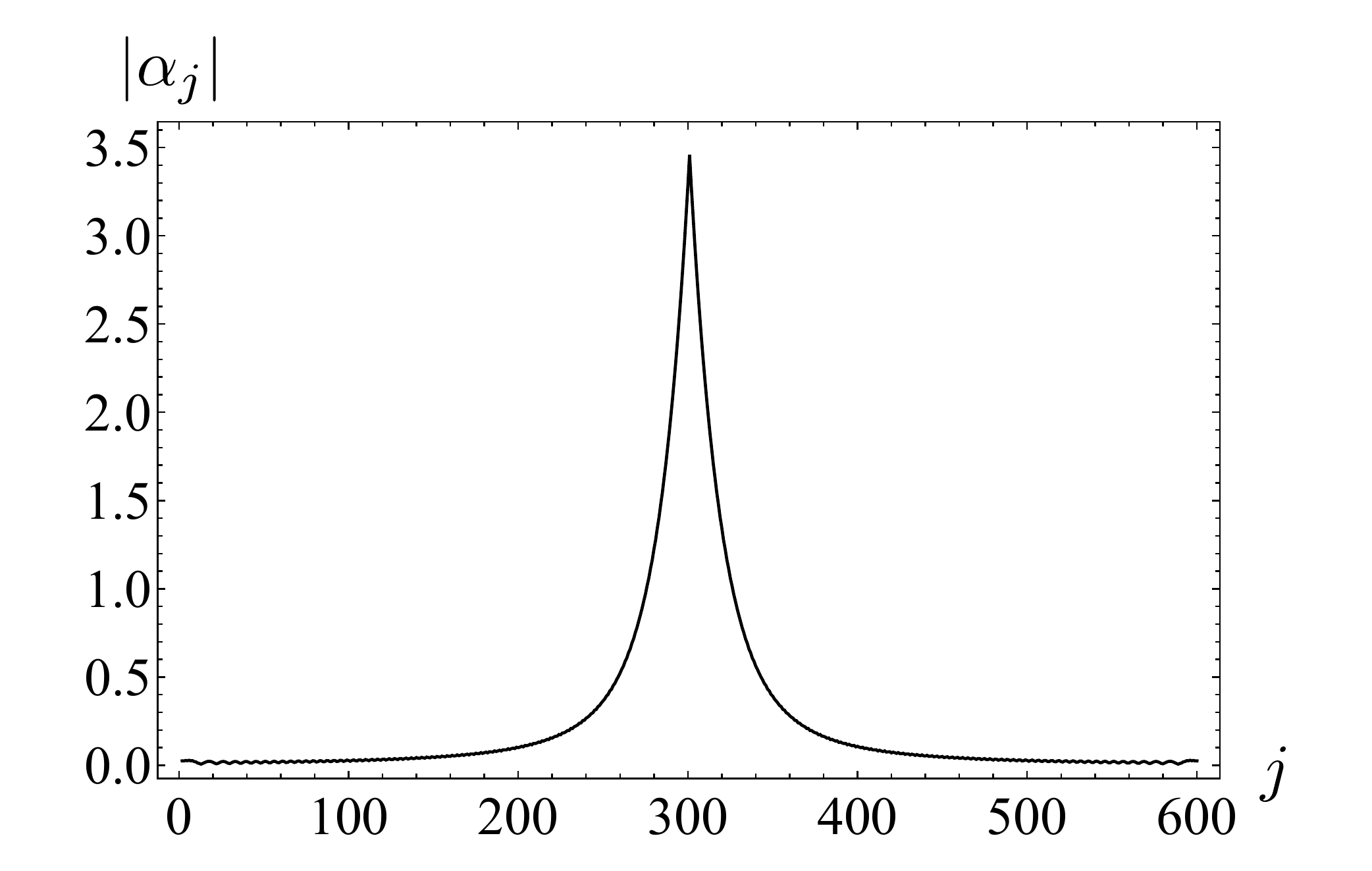}
     \caption{Panel {\bf (a)}: The amplitude $ \vert\alpha_j\vert $ of the perturbation induced in the harmonic environment due to the system $S$ for $ \omega=1$ (in units of $J$) and . The perturbation spreads through all the environmental sub-systems, which in turns determines a Markovian dynamics of the system only. This appears to be correlated to the emergence of quantum Darwinism in light of the redundant encoding of information entailed by the almost uniform spreading of information across ${\cal E}$. Panel {\bf (b)}: Same as panel {\bf (a)} but for $\omega/J=2.25 $, which sets the emergence of non-Markovian features due to the localisation of the perturbation over the array of harmonic oscillators. As we have discussed, this feature is correlated to the break-down of quantum Darwinism in light of the loss of redundancy in the information encoding process.}
     \label{ecc}
\end{figure*}

\begin{figure*}[t]
{\bf (a)}\hskip7cm{\bf (b)}
\includegraphics[scale=0.28]{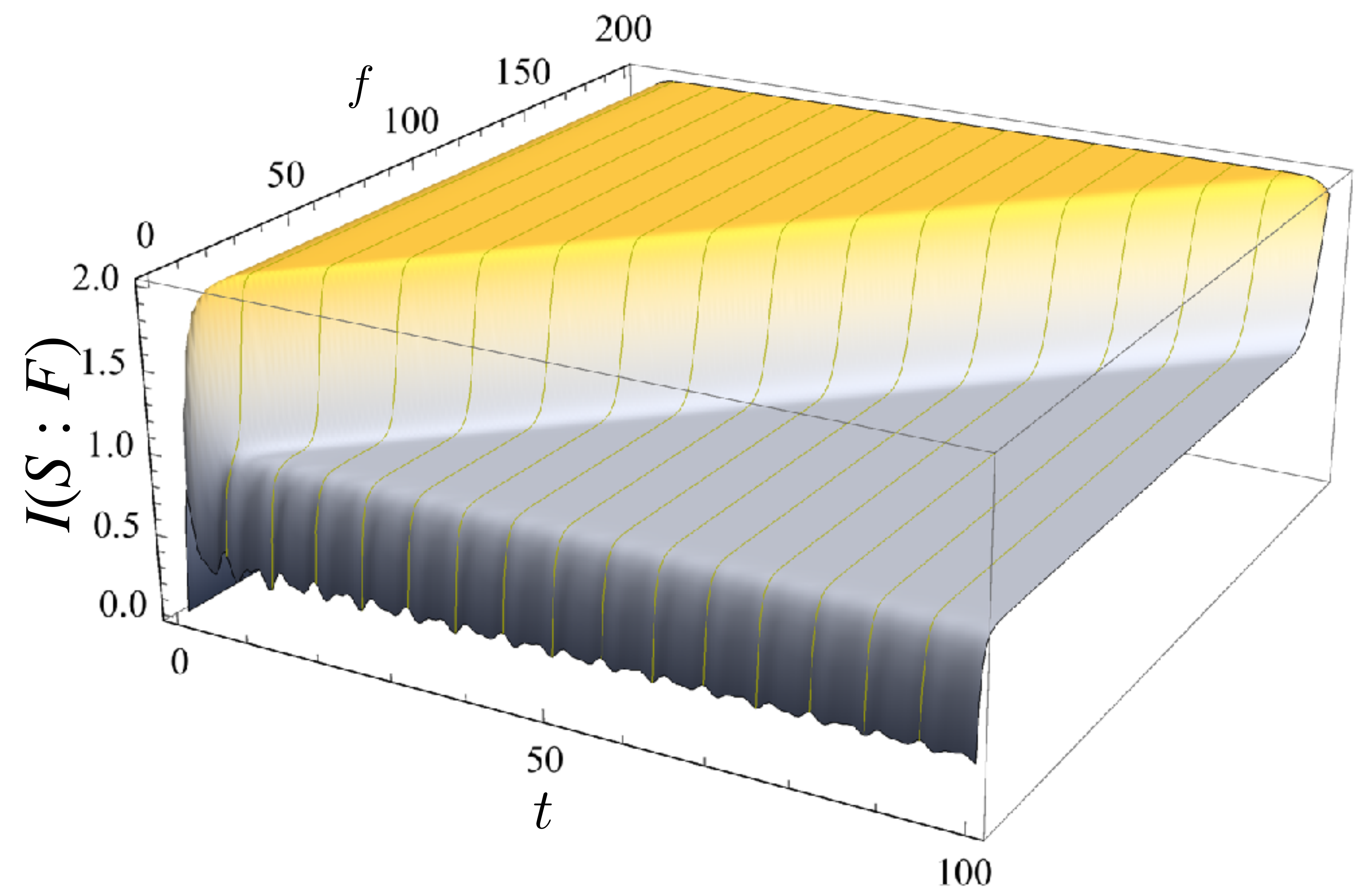}\includegraphics[scale=0.28]{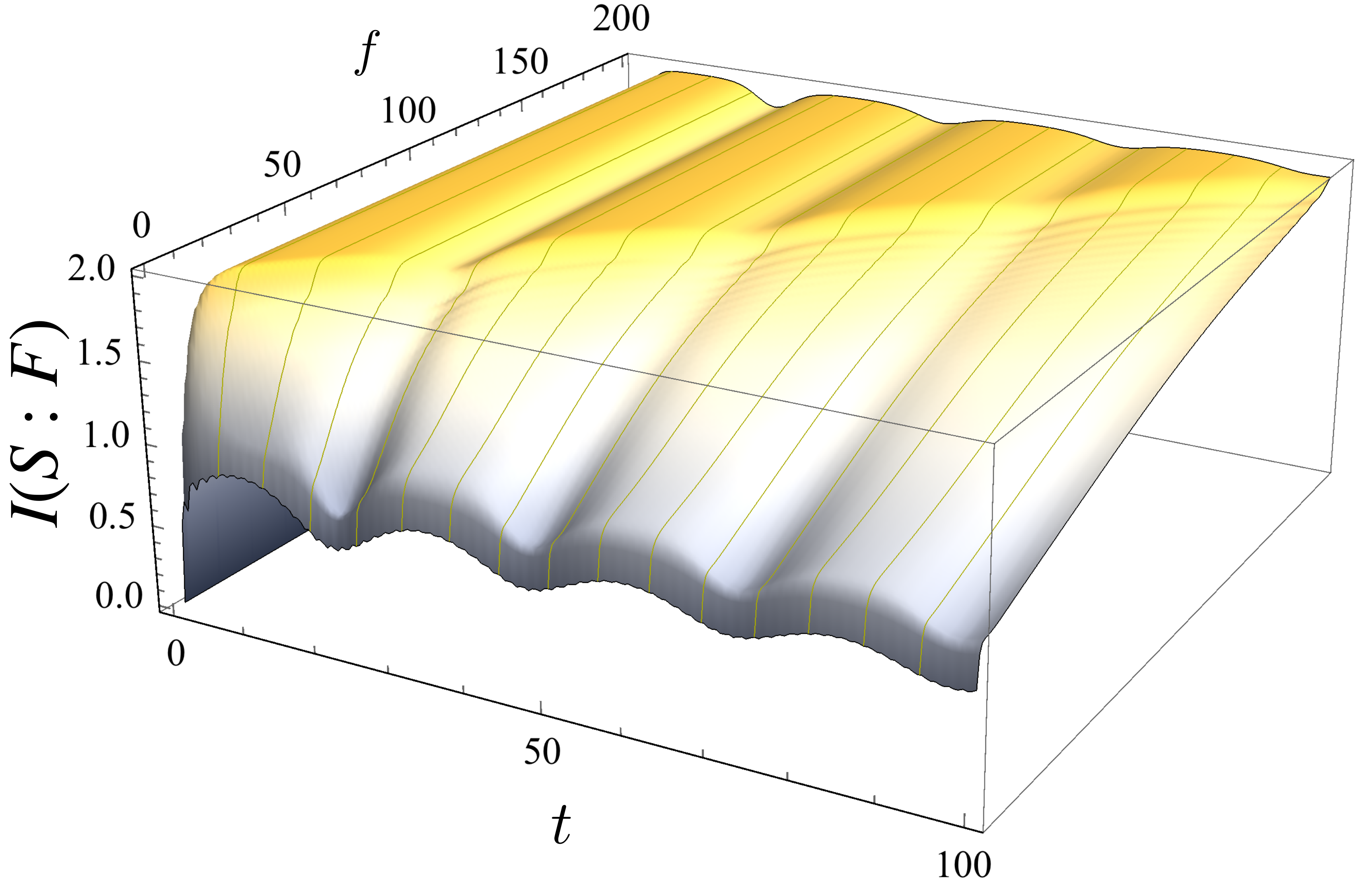}
\caption{Panel {\bf (a)}: Temporal PIP for $ \omega/J=0.5 $ and increasing size $f$ of the environmental fragment $F$ (total number of elements in ${\cal E}$ is $ N=201 $). As the time increases, quantum Darwinistic behaviours become apparent. In panel {\bf (b)}, which corresponds to $\omega/J=2.25$, Darwinistic features vanish, as the environment does not store information about \textit{S} redundantly. As \textit{f} increases, the amount of the information in the environment grows.}
\label{om05}
\end{figure*}

The correlation between Markovianity and Darwinism in our model can be understood in terms of information flow along the array of interacting harmonic oscillators. 
Let us first notice that, at $t=0$, $S$ induces a local perturbation on the array by displacing the central oscillator with $j=0$ from its equilibrium position. Due to the intra-environment coupling, such local perturbation propagates along the array. In Fig.~\ref{ecc} {\bf (a)} we show the amplitude $|\alpha_j|$ of the perturbation of the array sites again  for  interaction times long enough for the perturbation to reach the boundaries of the array (in the markovian regime) and $ \omega = J $, a working point that is associated with a Darwinistic and Markovian regime. Panel {\bf (b)} reports on the results valid for $\omega =2J$, which puts the system at the onset of non-Darwinistic and non-Markovian conditions. The analysis summarized in Fig.~\ref{ecc} thus shows clearly how the emergence of objective reality as witnessed by a Darwinistic phenomenology appears to be correlated to the features of information spreading across the environment. Working conditions giving rise to a {\it de facto} uniform spread of information across $F$ are associated with Darwinistic trends of $I(S:F)$ in light of the substantial redundance of information encoding about $S$. In turn, a widespread involvement of the set of environmental normal modes is bound to give rise to standard Markovian decoherence of the system's state. The situation becomes strikingly different when only a small part of the environment is affected by the local coupling to $S$. The finiteness of the effective environment gives rise to non-Markovian features as due to information trapping. In turns, Darwinism is prevented by the effective cut-off in the number of sub-environments involved in the ope dynamics of $S$. 
The temporal emergence of the features illustrated above is captured very well by an analysis that moves away from quasi stationary-state conditions and comprises a range of interaction times. Fig.~\ref{om05} reports on {\it dynamical} PIPs contrasting Markovian and non-Markovian regimes. In the Darwinistic/Markovian regime [cf. Fig.~\ref{om05}] as the perturbation propagates along the array the redundancy plateau $I(Sf) =1$ extends to larger fragment sizes at increasing interaction times before abruptly reaching its maximum value $I(S: F) =2$. 
When $ \omega/J>2 $, the encoding of information on \textit{S} in the environment is no longer redundant but rather increases for larger fragment sizes.

\section{Conclusions}
\label{conc}

Quantum Darwinism has been proposed as a framework for the characterization of the quantum-to-classical transition setting its premises in the objectivity of classical information about a system undergoing open quantum dynamics. Despite its appeal, the Darwinistic phenomenology is only partially understood so far, in particular in relation to the trade-off that its emergence sets with the rich dynamics of an open-quantum system. In order to advance our understanding of the phenomenology of quantum Darwinism, we have addressed the case of an exactly solvable interacting environmental model coupled locally to a single two-level system to give rise to a time-dependent dephasing dynamics. Our analysis illustrates a strong correlation between information trapping or spreading, which determine the degree of Markovianity of the ensuing system dynamics, and the emergence of Darwinistic behaviour in the amount of information that the environment acquires on the system itself. While providing a clear physical picture of the features of such important characteristics of the system's evolution, our study suggests a possible significant causal link between information spreading and the emergence of objective reality. 

\acknowledgments

This work was supported by the European Union through project TEQ (grant number 766900), the SFI-DfE Investigator Programme QuNaNet (grant number 15/IA/2864), and The Leverhulme Trust Research Project Grant. NM was partially supported by the Erasmus+ programme, and thanks the Centre for Theoretical Atomic, Molecular, and Optical Physics, School of Mathematics and Physics, Queen's University Belfast for hospitality during the early stages of this work.

\end{document}